\documentclass[fleqn,usenatbib]{mnras}

\usepackage{newtxtext,newtxmath}
\usepackage[T1]{fontenc}

\DeclareRobustCommand{\VAN}[3]{#2}
\let\VANthebibliography\thebibliography
\def\thebibliography{\DeclareRobustCommand{\VAN}[3]{##3}\VANthebibliography}

\usepackage{graphicx}	% Including figure files
\usepackage{amsmath}	% Advanced maths commands
\usepackage{placeins} % Ensures proper float placement
\usepackage{CJK}
\usepackage{scalerel}
\usepackage{soul}
\usepackage{gensymb}
\usepackage{dashrule}
\usepackage{tikz,siunitx}
\usepackage{fancyvrb}
\usepackage{fontawesome}
\usepackage{fontenc}
\usepackage{multirow}

\definecolor{linkcolor}{cmyk}{1,1,0,0}

\definecolor{mpltab:blue}{rgb}{0.1216,0.4667,0.7059}
\definecolor{mpltab:orange}{rgb}{1,0.5,0.05}
\definecolor{mpltab:green}{rgb}{0.173,0.627,0.173}
\definecolor{mpltab:red}{rgb}{0.84,0.15,0.16}
\definecolor{mpltab:purple}{rgb}{0.58,0.404,0.741}

\usetikzlibrary{shapes.geometric,shapes.symbols}

\newcommand{\tikzsymbol}[2][circle]{\tikz[baseline=-0.5ex]\node[inner
sep=2pt,shape=#1,draw,#2]{};}

\newcommand{\indep}{\perp \!\!\! \perp}
\newcommand{\dep}{\not\!\perp\!\!\!\perp}
\newcommand{\node}[1]{\mathsf{#1}}
\newcommand{\set}[1]{\mathbf{#1}}

\newcommand{\graph}{\mathcal{G}}

\newcommand{\parents}{\text{Pa}_{\graph}}

\newcommand\psj{PSJ} % The Planetary Science Journal

\title[TNO colours: formation location causes colour]{Causal evidence for the primordiality of colours in trans-Neptunian objects}

\author[Davis et al.]{
Benjamin L.\ Davis,$^{1,\dagger}$\thanks{E-mail: \url{ben.davis@nyu.edu} (BLD)}
Mohamad Ali-Dib,$^{1,\dagger}$
Yujia Zheng,$^{2,\dagger}$
Zehao Jin,$^{1,3,\dagger}$
Kun Zhang,$^{2,4}$
\newauthor and Andrea Valerio Macci\`{o}$^{1}$
\\
$^{1}$Center for Astrophysics and Space Science (CASS), New York University Abu Dhabi, PO Box 129188, Abu Dhabi, UAE\\
$^{2}$Carnegie Mellon University, Pittsburgh, PA, USA\\
$^{3}$Center for Astronomy and Astrophysics and Department of Physics, Fudan University, Shanghai 200438, People’s Republic of China\\
$^{4}$Mohamed bin Zayed University of Artificial Intelligence, Abu Dhabi, UAE\\
$^{\dagger}$These authors contributed equally to this work and are listed alphabetically.
}

\date{Accepted 2025 August 08. Received 2025 August 08; in original form 2025 June 15}

\pubyear{\the\year{}}

\begin{document}
\label{firstpage}
\pagerange{\pageref{firstpage}--\pageref{lastpage}}
\maketitle

% Abstract of the paper
\begin{abstract}
%This is a simple template for authors to write new MNRAS papers.
%The abstract should briefly describe the aims, methods, and main results of the paper.
%It should be a single paragraph not more than 250 words (200 words for Letters).
%No references should appear in the abstract.
The origins of the colours of Trans-Neptunian Objects (TNOs) represent a crucial unresolved question, central to understanding the history of our Solar System.
Recent observational surveys revealed correlations between the eccentricity and inclination of TNOs, and their colours.
This rekindled the long-standing debate on whether these colours reflect the conditions of TNO formation or their subsequent evolution.
We address this question using a model-agnostic, data-driven approach that unanimously converges to a common causal graph from the analysis of two different datasets, each from two different conditional independence test methods.
For evaluation, we demonstrate how our model is consistent with the currently-accepted paradigms of TNOs' dynamical histories, without involving any orbital modelling or physics-based assumptions.
Our causal model (with no knowledge of the existence of Neptune) predicts the need for an unknown confounding variable, consistent with Neptune's effects.
The model predicts that the colour of TNOs is the root cause of their inclination distribution, rather than the other way around.
This strongly suggests that the colours of TNOs reflect an underlying dynamical property, most likely their formation location.
Our model excludes formation scenarios that invoke substantial colour modification by subsequent evolution.
We conclude that the colours of TNOs are predominantly primordial.
\end{abstract}
% 200/200 words
% Select between one and six entries from the list of approved keywords.
% Don't make up new ones.

%%%%%%%%%%%%%%%%%%%%%%%%%%%%%%%%%%%%%%%%%%%%%%%%%%

%%%%%%%%%%%%%%%%% BODY OF PAPER %%%%%%%%%%%%%%%%%%

\section{Introduction}\label{sec:intro}

Trans-Neptunian Objects (TNOs) are invaluable probes into the history and evolution of our Solar System \citep{morby1}. 
However, the wealth of information they encode is difficult to decipher with numerous, often opposing interpretations of the their properties.
This includes intrinsic characteristics such as their sizes and correlated properties such as their orbits and surface photometric colours.
The last two have long been closely examined in an effort to unravel the relation between them \citep{Jewitt:2001,liuip2019,2022ApJ...937L..22C,berdar2025}.

There are three leading theories regarding the origin of TNO colours: 
\begin{enumerate}
\item The primordial origin hypothesis of the TNO colour diversity argues that TNO colours reflect compositional gradients in the protoplanetary disk, preserved since formation \citep{brown2012,nesvorny2020,Ali-Dib:2021,buchanan,alonso2025}.
Objects formed at different heliocentric distances thus acquired distinct volatile and refractory compositions, leading to colour variations.
For example, objects that formed beyond the CO and N$_2$ snowlines could have acquired redder surfaces.
Dynamical processes (e.g., planetary migration and scattering) later redistributed these bodies into their current orbits, imprinting correlations between colour and orbital parameters like inclination.
In this scenario, using causality theory jargon, inclination (inc) is said to be caused by the colours, which is indicative of the formation location.
In reality, an additional parameter, $a_\mathrm{ini}$ (the formation location), sets the inclination and colour simultaneously.
The initial semimajor axis thus implies something directly about colour, and the scattering of TNOs by Neptune causes $a_\mathrm{ini}$ to be related to the present $e$ and inc.
\item However, alternatively, many works \citep{Luu:1996b,Stern:2002,Hainaut2002,Doressoundiram2003,2018MNRAS.481.1848A} argued that collisional evolution is the origin of TNO colours, where collisions expose fresh subsurface ices or organic materials, altering albedo and spectral slopes. 
Dynamically excited populations (higher $e$ and inc) experience more frequent collisions due to orbital crossings, leading to colour--inclination correlations.
This framework treats colour as a secondary property shaped by post-formation bombardment.
Opponents of this model argue that if collisional resurfacing were causal, dynamically excited populations would exhibit homogenised colours over time due to frequent mixing.
\item A third possibility proposed that initially diverse bulk compositions undergo selective volatile evaporation post-formation, establishing steep compositional gradients across the primordial disk that, coupled with subsequent UV photolysis and particle irradiation, yield distinct surface chemistries.
A key difference between this and the `primordial origin' hypothesis is the necessity of post-formation irradiation, either pre-instability \citep{brown1,wongbrown1,wongbrown} or post-instability \citep{Zuzana}.
From a causality lens, a post-instability irradiation model introduces a causal relationship between the current semimajor axis and the colour of TNOs.
\end{enumerate}
All of the three theories explain the observed correlation between colour and inclination, therefore the origin of TNO colours remains a long-standing debate.
However, the three models do not share the same causation: the first primordial origin model implies that colour causes the current inclination; the second collisional model, in contrary, demands inclination to cause colour; while the third post-formation theory requires not only colour to cause inclination, but also the current semimajor axis to cause colour.
If one can find the causal structure between TNO colours and their orbital parameters, the three theories will be distinguishable.

Identifying cause-effect relationships is crucial for moving beyond mere correlation to uncover the underlying causal mechanisms governing a system.
Traditionally, causal relationships are established through interventions or randomised experiments, where one variable is explicitly manipulated while all others are held constant, and the resulting effects are observed.
However, such interventions are infeasible in fields like astronomy, where the `test subjects' exist at unreachable astronomical distances.
Consequently, advanced methods are required to infer causal relationships from purely observational data---an endeavour that lies at the core of causal discovery \citep{spirtes2000causation}.

For decades, causal discovery has been a transformative tool in science, enabling researchers to look beyond correlation and uncover the fundamental mechanisms driving complex systems.
Its applications in biology are extensive, from mapping intricate protein signalling pathways \citep{friedman2004inferring} to deciphering gene regulatory networks \citep{sachs2005causal}.
The methodology's utility extends into physics, where it is invaluable for analysing systems that defy direct experimentation; examples include identifying the drivers of plasma instabilities in fusion reactors and modelling emergent causal structures within condensed matter \citep{runge2019inferring}.
Although a newer frontier for astrophysics, recent works are beginning to demonstrate the power of causal discovery in decoding astronomical data \citep{Pasquato:2023,Pasquato:2024,Jin:2024,Jin:2025AAS,Jin:2025,Davis:2025}.
From cellular processes to cosmic structures, this approach provides a robust framework for modelling the underlying causal architecture of the natural world, built upon the foundational contributions of \citet{spirtes2000causation} and \citet{pearl2009causality}.

In this letter, we use a purely data-driven, model-agnostic, statistical causal discovery method to study the causal structure among the dynamical parameters and colours of TNOs, revealing a causal structure consistent with model one, the primordial origin model, while ruling out the other two.
We show that not only this technique allows us to derive some of the main lines of the current consensus on the origins of TNOs, but also that it elucidates the direction of causality between the dynamical parameters and colours of TNOs, and predicts the existence of an unknown perturbing body, i.e., Neptune.
We first give an overview of our data sample (\S\ref{sec:data}), detail our causal discovery methods (\S\ref{sec:methods}), present the results of our analysis (\S\ref{sec:results}), and finally conclude with an overall summary (\S\ref{sec:discussion}).

\section{Data}\label{sec:data}

\begin{figure*}
\centering
\includegraphics[clip=true, trim= 2mm 3mm 3mm 3mm, width=\linewidth]{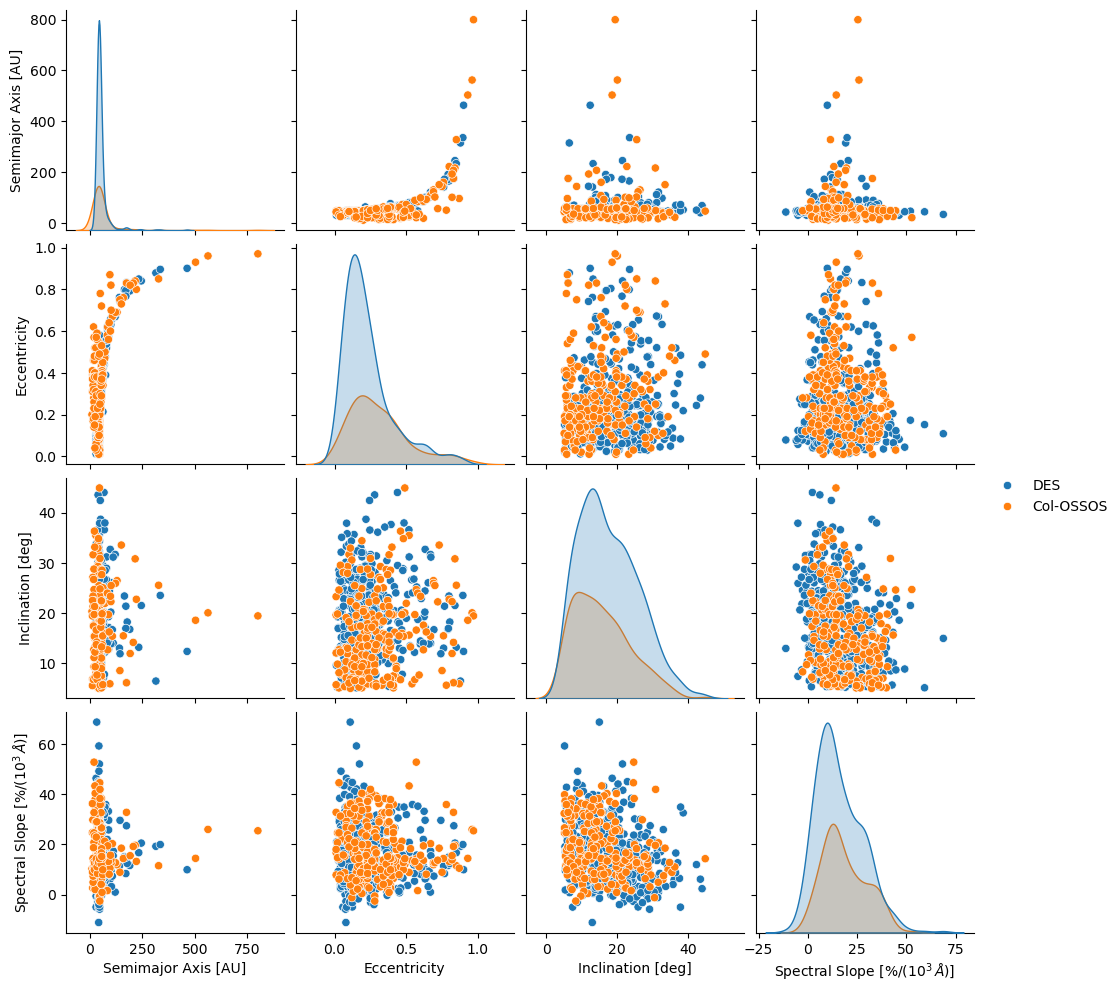}
\caption{
Pairplot of all 229 and 674 TNOs in the Col-OSSOS and DES datasets, respectively.
The colour--eccentricity and colour--inclination correlations hold in both datasets.
}
\label{fig:pairplot}
\end{figure*}

We use two separate datasets for this work: the Col-OSSOS survey, and Dark Energy Survey (DES). Both include hot classicals, centaurs, and resonant/scattered objects.
For each TNO (see Fig.~\ref{fig:pairplot}), we have three orbital elements: semimajor axis ($a$), eccentricity ($e$), and inclination (inc); and we have spectral slope (i.e., colour).
A fundamental assumption of this work is that colours are primordial, and thus strongly correlated to the initial location of a TNO.
Hereafter, we treat colours as a proxy for the initial semimajor axis of the objects. 
In the following, the two datasets are analysed separately as they originate from different surveys with different characteristics and observational biases.

\subsection{The Colours of the Outer Solar System Origins Survey}

Our first dataset is based on (but not exclusively) the Colours of the Outer Solar System Origins Survey \cite[Col-OSSOS;][]{Schwamb:2019}.
It was originally assembled by \citet{Marsset:2019} and then re-used by \citet{Ali-Dib:2021}.
It contains a total of 229 TNOs in a dataset for which discovery biases were modelled.
These consist of Hot Classicals (48), Resonant (102), Centaur (36), Scattered (28), and Detached (15) objects. 

This sample shows a bimodal distribution of optical spectral slopes ($s$); a Gaussian mixture model fit to the histogram indicated that the two colour classes intersect at \(s \simeq 20.6\%\,(10^3\AA)^{-1}\), corresponding to \((g{-}r) \approx 0.78\) and \((V{-}R) \approx 0.56\).
This value agrees with thresholds adopted in earlier works \citep[see][Table~2]{Marsset:2019} and provides a boundary between less‑red and very‑red objects: TNOs with \(s + \delta s < 20.6\%\,(10^3\AA)^{-1}\) were labelled `gray/LROs,' (i.e., less-red objects) those with \(s - \delta s > 20.6\%\,(10^3\AA)^{-1}\) were labelled `very‑red/VROs,' (i.e., very-red objects) and objects whose uncertainties straddle the boundary were left unclassified and out of the dataset.
Using this classification, \citet{Marsset:2019} found that the very‑red population has a cut‑off inclination of $\sim$$21^\circ$, whereas gray objects extend to higher inclinations.
Their results were subsequently expanded by \citet{Ali-Dib:2021}, who reported a similar cut‑off in eccentricity around \(e \approx 0.42\) for the VROs.
\citet{Ali-Dib:2021} concluded that there is a paucity of VROs in the scattered disk and used a Solar System formation model to explain these trends as a consequence of their formation location in the disk. 

\begin{figure*}
    \centering
    \includegraphics[clip=true, trim= 3mm 2mm 2mm 2mm, width=\linewidth]{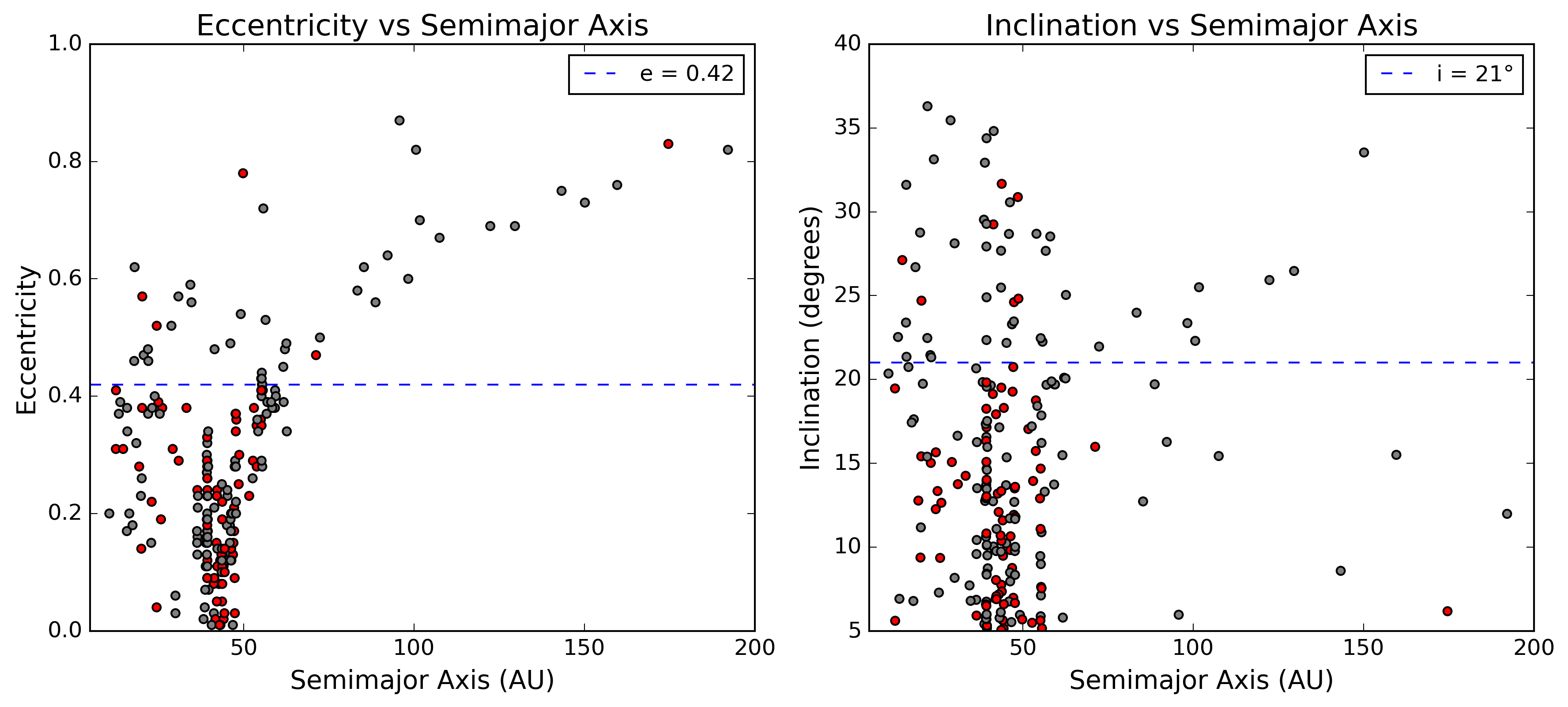}
    \includegraphics[clip=true, trim= 3mm 2mm 2mm 2mm, width=\linewidth]{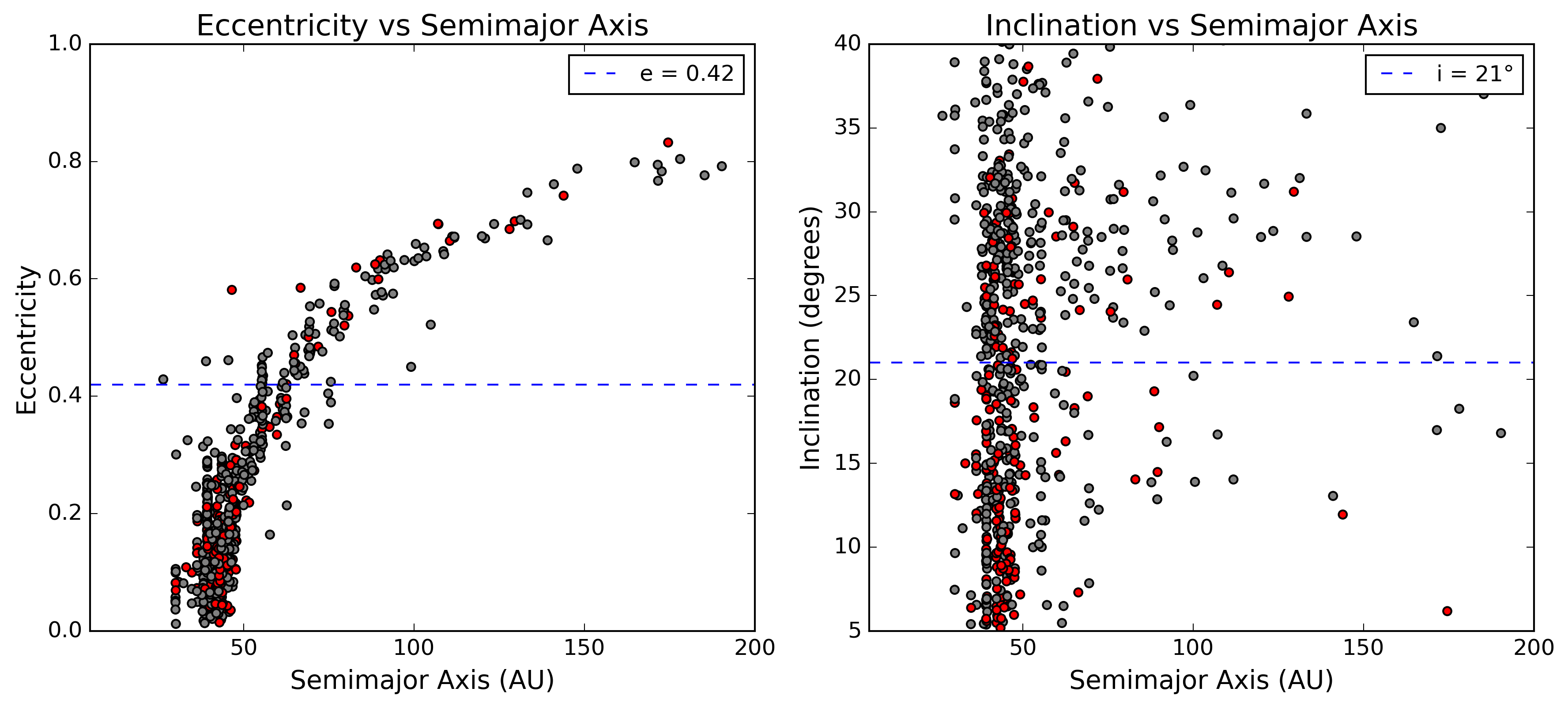}
    \caption{
    The \citet{Marsset:2019} and \citet{Ali-Dib:2021} (top), and DES (bottom) samples shown as $a$--$e$ (left) and $a$--inc (right) plots.
    Colours were defined such that red (\tikzsymbol{fill=red}) is for Very Red Objects (spectral slopes higher than 20.6\%/(10$^3$\AA)) and gray (\tikzsymbol{fill=gray}) is for Less Red Objects.
    The plots clearly show the paucity of VROs for eccentricities higher than 0.42 and inclinations higher than 21$\degree$, respectively (\textcolor{blue}{\hdashrule[0.35ex]{8mm}{1pt}{1mm}}).
    }
    \label{fig:dataset}
\end{figure*}

Our dataset is further summarised in Fig.~\ref{fig:dataset}.
We define VROs as TNOs with spectral slopes greater than 20.6\%/(10$^3$\AA).
The colour--eccentricity correlation is revealed in this plot as a paucity of VROs for eccentricity above 0.42.
Similarly, the colour--inclination correlation manifests itself as a lack of VROs for inclinations above 21$\degree$.

\subsection{The Dark Energy Survey}

The Dark Energy Survey \citep{Bernardinellia,berdar2025} sample consists of 814 TNOs with absolute magnitude $5.5 < H_r < 8.2$ and accurate colours. 
The obtained optical colour classes were found to be an identical match to the different compositional classes derived from \textit{JWST} IR spectra (when available). 
For self-consistency between the two datasets, we removed cold objects with inclinations below 5$\degree$, and very far objects beyond the maximal semimajor axis of the Col-OSSOS dataset. 
We thus end up with 674 TNOs spanning most dynamical classes: 274 Hot Classicals, 209 Resonant, 145 Detached, 45 Scattered, and one Centaur.\footnote{Note that centaurs are much less prominent than in the Col-OSSOS sample.
Thus, the more unbalanced DES sample is less suited for demographic studies between the subpopulations.}
Our final datasets are shown in Fig.~\ref{fig:pairplot}, where the correlations between colour and inclination, and colour and eccentricity are noticeable in both samples, confirming that they are not merely an observational bias in the Col-OSSOS sample.\footnote{Note that \cite{berdar2025}, used this survey to find a paucity of `near-infrared faint' SDOs (scattered-disk objects) in the data.}

\section{Methodology: causal discovery}\label{sec:methods}

The foundation of causal discovery lies in uncovering the footprints of causality embedded in data.
One of the most important sources of such information is conditional independency (CI) relations.
In this section, we will first show how different causal structures leaves different CI footprints, then explain how to utilise these CIs to constrain causal structures even with the presence of latent variables, and finally we introduce the Fast Causal Inference (FCI) algorithm used in this work.
For further reading on causal discovery and causality, see \textit{Causation, Prediction, and Search} \citep{spirtes2000causation}, \textit{Causality} \citep{pearl2009causality}, or a review for astrophysicists in \citet[][\S2]{Jin:2025}.

\subsection{Conditional independency footprints}

\begin{figure}
  \centering
  \includegraphics[width=\linewidth, trim= 8mm 6mm 21mm 2mm, clip=true]{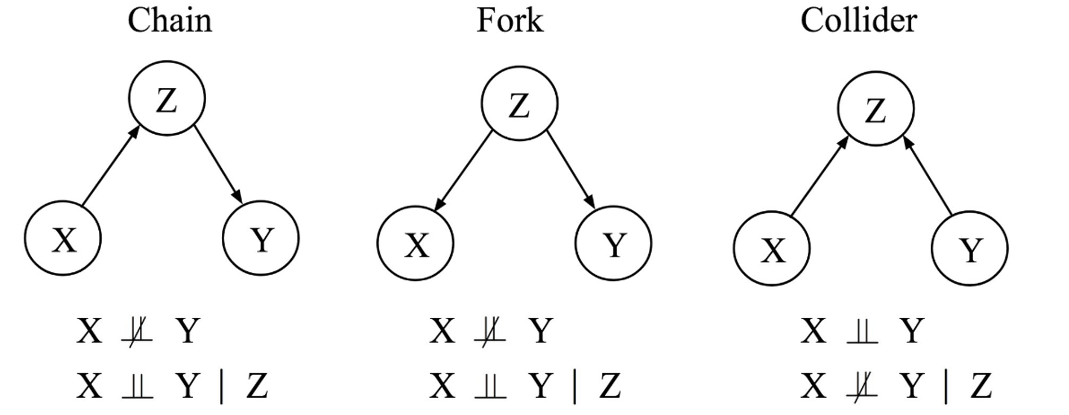}
  \caption{
  Three basic causal blocks, and their resulting set of (conditional) independencies. $\dep$ denotes dependent, $\indep$ means independent, and $|$ is the notation for condition.
  For example, `$X \indep Y \ |\ Z$' means `$X$ is independent to $Y$ when conditioned on $Z$.'}
\label{fig:chain_fork_collider}
\end{figure}

The causal structure among a set of variables is ideally represented by a Directed Acyclic Graph (DAG), consisting of nodes and directed edges (i.e., arrows), where directed edges between nodes denote the direction of causality.
There are three basic causal structures: \emph{chains}, \emph{forks}, and \emph{colliders} (Figure~\ref{fig:chain_fork_collider}), each of them carrying two (conditional) independency signatures. 
These sets of CIs naturally hold given each DAG, as long as a rather general assumption called the Markov assumption\footnote{The Markov assumption states that given its parents in a DAG, a node is independent of all its non-descendants.} holds.
It is worth noting that although a chain and a fork share the same CIs, a collider carries a different set of CIs, making it possible to constrain causal structures.
The three basic causal structure serve as the building block towards more composite DAGs with more than three variables.
The CIs in these larger DAGs are determined by all the paths among variables, where each path is made of chains, forks, or colliders.
The CIs of chains, forks, and colliders propagate along the paths following separation rules, which we will not go into detail here.

\subsection{Causal discovery with latent variables}

By analysing conditional independencies among different components of an observed system, we can infer causal relationships between pairs of variables.
This allows us to construct a graph that encodes the results of essential conditional independence tests, revealing which variables \emph{cause} others under appropriate conditions.
Ideally, the output is a DAG for a unique solution (for example, in the case of a collider) or a Completed Partially Directed Acyclic Graph (CPDAG) for a Markov equivalence class, where the direction of some of the edges cannot be determined (for example, in the case of a chain or a fork).

However, since it is impossible to measure all variables in the Universe, latent variables are always present.
These unmeasured variables can significantly impact the correctness of the causal structure discovered.
For example, suppose that $X$ and $Y$ are independent in the general population, but a sample is selected based on a variable $Z$ that influences both $X$ and $Y$.
In that case, $X$ and $Y$ may exhibit statistical dependence in the sample, even though no such relationship exists in the population.
This can lead to spurious causal conclusions, falsely suggesting a direct causal relationship between $X$ and $Y$.

To address this challenge, we employ a principled approach capable of uncovering causal relationships even in the presence of latent variables.
A widely used method for this purpose is Fast Causal Inference \citep[FCI;][]{spirtes1995causal, zhang2008completeness}, a constraint-based algorithm that has been proven to provide sound causal conclusions despite unmeasured variables.
FCI has been applied across various scientific domains, including biology, economics, and climate science.
For our analysis of TNO orbital elements and colours, we use the FCI implementation in the \texttt{Python} package \texttt{causal-learn} \citep{zheng2024causal} to infer the underlying causal structure.

\subsection{The Fast Causal Inference algorithm}

Unlike many causal discovery methods that assume that all relevant variables are measured (such as those producing DAGs or CPDAGs), FCI accounts for the possibility of unobserved variables.
As a result, its output is a Partial Ancestral Graph (PAG), which provides more nuanced causal information.

\subsubsection{Partial Ancestral Graph notations}\label{sec:pag_notations}

In a PAG, a bi-directional arrow $X\longleftrightarrow Y$ corresponds to a confounding relation (i.e., a third variable causes both X and Y) and empty circles ($\circ$) represent uncertainty regarding the ending symbol of an edge.
Specifically, the edges in a PAG have the following interpretations:
\begin{itemize}
    \item $X \xrightarrow{\hspace{4.5mm}} Y $: $X$ is a \textit{cause} of $Y$.
    \item $X\longleftrightarrow Y$: There is a latent common cause of $X$ and $Y$.
    \item $X\ \circ\hspace{-1.78mm}\longrightarrow Y$: $Y$ is not an \textit{ancestor} of $X$, i.e., either $X\rightarrow Y$ or $X\longleftrightarrow Y$, but not $X\leftarrow Y$.
    \item $X\ \circ$---$\circ \  Y$: No set $d$-separates $X$ and $Y$, i.e., $X\rightarrow Y$, $X\leftarrow Y$, and $X\longleftrightarrow Y$ are all possible.
\end{itemize}
By introducing these notations, we can account for latent variables in the discovery process and uncover causal relations among measured variables while acknowledging uncertainties introduced by unmeasured factors.
More importantly, when the algorithm cannot determine a definitive causal direction due to latent variables, it explicitly represents this uncertainty rather than arbitrarily assigning a direction.
This principled approach distinguishes causal analysis from correlation-based techniques, ensuring that conclusions are drawn with a clear acknowledgment of underlying assumptions and limitations.

\subsubsection{FCI: A two-stage algorithm based on CI tests}\label{sec:fci_steps}

FCI is a two-stage algorithm to discovers causal relationships based on a series of CI tests.
The idea is to start with a fully-connected graph\footnote{In a fully-connected graph, every node is connected to every other node with the most general type of edge $\circ$---$\circ$, allowing any type of potential causal connection between any pair of nodes.}, where all causal structures are allowed, and gradually constrain the graph when the CIs encoded by the graph are inconsistent with the CIs found in the data.
Specifically, the algorithm proceeds in the follow two stages:
\begin{enumerate}
    \item \emph{Skeleton Discovery}: Starting with a fully connected graph representing all possible causal structures, FCI removes edges when two nodes are independent, and when two nodes become conditionally independent given some subset of other variables. 
    By the end of this stage, FCI arrives at an the undirected graph called a `skeleton.'
    \item \emph{Orientation}: The second stage is to orient the remaining edges. FCI orients the edges based on the separation information, collider detection (e.g., identifying V-structures), and propagation of orientation constraints. 
\end{enumerate}
We will walk readers through the above two FCI stages in our specific TNO case in \S\ref{sec:steps}.

\subsubsection{Conditional independence tests}\label{sec:CI_tests}

CI tests are performed along with the two stages of FCI (mostly during stage~i, and the results are re-used during stage~ii) whenever necessary.
Here, we adopt two commonly used CI tests in causal discovery, including the Fisher Z-test \citep{fisher1921probable} and the Kernel-based Conditional Independence (KCI) test \citep{zhang2011kernel}. 
\begin{itemize}
    \item The Fisher Z-test assesses conditional independence by measuring partial linear correlations between variables, providing a fast and effective method for detecting dependence relations.
    \item In contrast, the KCI test is a non-parametric approach based on reproducing kernel Hilbert space (RKHS) embeddings, allowing it to capture complex non-linear dependencies without assuming specific functional forms.
\end{itemize}
Like many other statistical tests, the output of the Fisher Z-test or the KCI test is a $p$-value which gives the possibility against a null (conditionally dependent) hypothesis. 
For example, if the CI test between $A$ and $B$ conditioned on both $C$ and $D$ has a $p$-value of 0.05, then $A \dep B\ |\ \{C,D\}$ at the 95\% confidence level.

While KCI offers greater flexibility and consistency in general settings, it is computationally more intensive and sensitive to kernel choices.
The selection of CI test depends on the trade-off between computational efficiency and the need for non-parametric estimation. 
Here, we perform both the Fisher Z-test and the KCI test with polynomial kernels to both Col-OSSOS and DES datasets.
The CIs and $p$-values are reported in Table~\ref{tab:CI}.
For any dataset and any CI test (i.e., any of the four columns), $a \indep \text{inc}$, $a \indep \text{colour}$, and $e \indep \text{colour}$ show significant higher $p$-values compared to others within the same column, indicating they are independent, while other CIs listed are dependent.
Quantitatively, the CIs are determined at least above the 98.2\% confidence level for Col-OSSOS with the Fisher Z-test, 91.5\% for Col-OSSOS with the KCI test, 99.9\% for DES with the Fisher Z-test, and 99.7\% for DES with the KCI test.\footnote{There are more data points in DES (674) than in Col-OSSOS (229), therefore the CI tests on DES data show lower $p$-values (i.e., higher confidence levels) than the ones on Col-OSSOS data, as expected.}

\begin{table*}
\centering
\caption{
List of conditional independencies, type of conditional independence tests performed, and the $p$-value of the statistical test.
$\dep$ denotes dependent, $\indep$ means independent, and $|$ is the notation for condition.
The Fisher-Z and KCI tests are performed on both Col-OSSOS and DES data.
The $p$-value for the null hypothesis for each CI test is shown.
A $p$-value closer to zero suggests dependence, and a $p$-value closer to one favours independence.
Both tests on both datasets unanimously show that $a \indep \text{inc}$, $a \indep \text{colour}$, and $e \indep \text{colour}$, while other CIs listed are dependent.
The PAG derived in Fig.~\ref{fig:steps} and later shown in Fig.~\ref{fig:fci} directly comes from this list of conditional independencies following the FCI algorithm detailed in \S\ref{sec:fci_steps} and \S\ref{sec:steps}.
}
\label{tab:CI}
\begin{tabular}{lcccc}
\hline
\multirow{3}{*}{Conditional independencies} & \multicolumn{4}{c}{Conditional independence test $p$-value}            \\
                                            & \multicolumn{2}{c}{Col-OSSOS} & \multicolumn{2}{c}{DES}      \\
                                            & Fisher-Z         & KCI        & Fisher-Z        & KCI        \\ \hline
$a \dep e$                                  & 0.000            & 0.000      & 0.000           & 0.000      \\ 
$a \indep \text{inc}$                       & 0.144            & 0.421      & 0.006           & 0.568      \\ 
$a \indep \text{colour}$                    & 0.918            & 0.488      & 0.677           & 0.250      \\ 
$e \dep \text{inc}$                         & 0.005            & 0.085      & 0.000           & 0.001      \\ 
$e \indep \text{colour}$                    & 0.211            & 0.135      & 0.181           & 0.010      \\ 
$\text{inc} \dep \text{colour}$             & 0.001            & 0.009      & 0.000           & 0.000      \\ 
$a \dep e\ |\  \text{inc}$                  & 0.000            & 0.068      & 0.000           & 0.001      \\ 
$e \dep \text{inc}\ |\ a$                   & 0.018            & 0.070      & 0.000           & 0.002      \\ 
$e \dep \text{inc}\ |\ \text{colour}$       & 0.010            & 0.041      & 0.000           & 0.003      \\ 
$\text{inc} \dep \text{colour}\ |\ e$       & 0.003            & 0.004      & 0.000           & 0.000      \\ 
\hline
\end{tabular}
\end{table*}

\subsubsection{Step-by-step derivation of the PAG}\label{sec:steps}

Here, we manually derive of the causal structure (i.e., the PAG) following the two stages of FCI outlined in \S\ref{sec:fci_steps} using the list of CIs discussed in the previous section (\S\ref{sec:CI_tests}) and shown in Table~\ref{tab:CI}.
The whole process is automated in the \texttt{Python} package \texttt{causal-learn} \citep{zheng2024causal}, but here we explicitly write it out for readers new to causal discovery.

\begin{figure*}
    \centering
    \includegraphics[width=\linewidth, trim={{0.03\linewidth} {0.23\linewidth} {0.27\linewidth} {0.19\linewidth}}, clip=true]{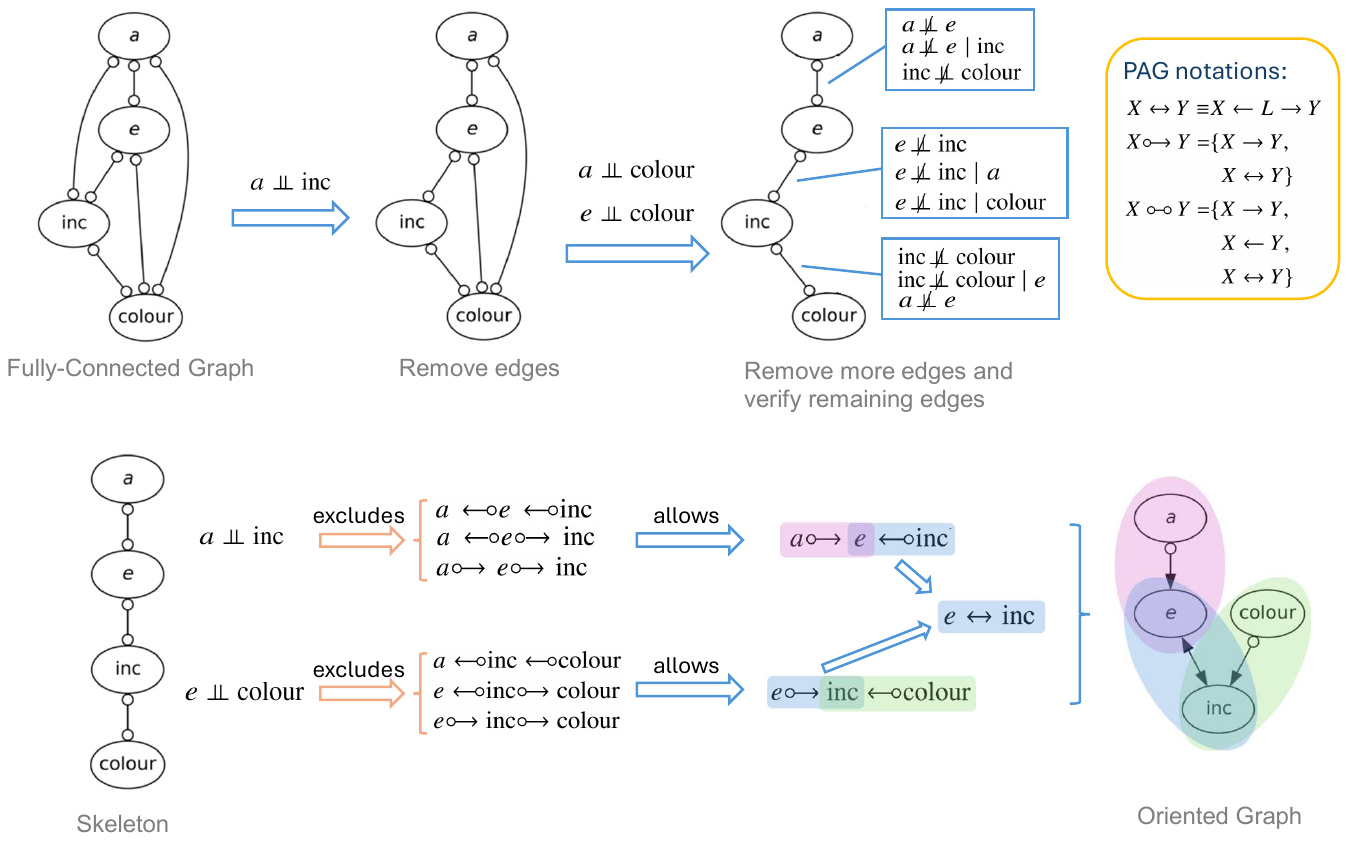}
    \caption{
    The visualisation of the FCI algorithm.
    In FCI stage~i (upper panel), the algorithm starts with a fully-connected graph where all nodes are interconnected with $\circ$---$\circ$ to allow all possible cases. 
    Then, the algorithm goes through every edge and removes an edge when two nodes are independent, and when two nodes become conditionally independent given some subset of other variables.
    In FCI stage~ii (lower panel), the remaining edges are oriented by constraining the ending symbols of each edge according to conditional independencies.
    }
    \label{fig:steps}
\end{figure*}

During FCI stage~i, we start from the hypothesised, undirected, and fully-connected graph in Fig.~\ref{fig:steps} (top panel) with our four nodes ($a$, $e$, inc, and colour), and remove edges between independent nodes.
For example, given $a \indep \text{inc}$, it is then not possible to have $a$ cause \text{inc} ($a\rightarrow\text{inc}$), \text{inc} cause $a$ ($\text{inc}\rightarrow a$), nor a third latent variable ($L$) cause both $a$ and inc ($a\leftarrow L \rightarrow \text{inc}$, shortened as $a\leftrightarrow\text{inc}$).
Therefore, the edge between $a$ and inc ($a\ \circ$---$\circ \  \text{inc}$) can be removed.
Similarly, $a\ \circ$---$\circ \  \text{colour}$ and $e\ \circ$---$\circ \  \text{colour}$ are also removed since $a \indep \text{colour}$ and $e \indep \text{colour}$.
The remaining edges are valid since the nodes are dependent with or without conditioning on other non-latent nodes.
For example, $e \dep \text{inc}$, $e \dep \text{inc}\ |\ a$, and $e \dep \text{inc}\ |\ \text{colour}$ together requires an edge between $e$ and inc.\footnote{$e \dep \text{inc}$ alone does not guarantee an edge between $e$ and inc, as $e\leftarrow a \rightarrow\text{inc}$ or $e\leftarrow \text{colour} \rightarrow\text{inc}$ can both lead to $e \dep \text{inc}$.
However $e \dep \text{inc}\ |\ a$ and $e \dep \text{inc}\ |\ \text{colour}$ rules out these two cases and secures an edge between $e$ and inc.}

Now, we are left with with a skeleton graph as Fig.~\ref{fig:steps} (bottom panel), and we can move to FCI stage~ii---orient the remaining edges according to CIs.
Given the current skeleton, $a \indep \text{colour}$, $a \indep \text{inc}$, and $e \indep \text{colour}$ together form a classical setup where there must be a latent confounder between $e$ and inc.
Consider $a$, $e$, and inc, both a chain structure (i.e., $a\circ\hspace{-1.7mm}\rightarrow e\circ\hspace{-1.7mm}\rightarrow\text{inc}$ or $a\leftarrow\hspace{-1.7mm}\circ e\leftarrow\hspace{-1.7mm}\circ\text{inc}$) and a fork structure (i.e., $a\leftarrow\hspace{-1.7mm}\circ e\circ\hspace{-1.7mm}\rightarrow\text{inc}$) are forbidden as they will not satisfy the fact that $a \indep \text{inc}$.
The only structure compatible with $a \indep \text{inc}$ is a collider (i.e., $a\circ\hspace{-1.3mm}\rightarrow e \leftarrow\hspace{-1.3mm}\circ\text{inc}$).
Similarly, we can find $e\circ\hspace{-1.3mm}\rightarrow \text{inc} \leftarrow\hspace{-1.3mm}\circ\text{colour}$ according to $e \indep \text{colour}$.
The need for both $e \leftarrow\hspace{-1.3mm}\circ\text{inc}$ and $e\circ\hspace{-1.3mm}\rightarrow \text{inc}$ calls for a latent confounder $L$ causing both $e$ and inc (i.e., $e\leftarrow L \rightarrow\text{inc}$, or $e\leftrightarrow \text{inc}$ in a more compact notation).
We therefore arrive at the final PAG in Fig.~\ref{fig:steps} (bottom panel) and later shown in Fig.~\ref{fig:fci}.

\subsubsection{Validation of FCI with generated data}

The FCI algorithm is a time-tested algorithm that has been proven successful both in idealised data \citep{colombo2012learning} and real-world data \citep{glymour2019review}.
Here, we perform a simple test with generated ideal data from latent linear Structural Causal Models (SCMs).
Such models can be defined as a DAG $\graph:=(\set{V}_\graph,\set{E}_\graph)$, where each variable $V_i \in \mathbf{V}_{\mathcal{G}}$ is generated following a latent linear SCM:
\begin{equation}
    \label{eq:lem}
\node{V}_i=\sum \nolimits_{\node{V}_j \in \parents(\node{V}_i)} a_{ij} \node{V}_j + \varepsilon_{\node{V}_i},
\end{equation}
where $\set{V}_\graph:=\set{L}_\graph\cup\set{X}_\graph$ contains a set of $n$ observed variables ($\set{X}_\graph:=\{\node{X}_i\}_{i=1}^{n}$) and $m$ latent variables ($\set{L}_\graph:=\{\node{L}_i\}_{i=1}^{m}$).
$\parents(V_i)$ is the parent set (i.e., nodes that directly cause $V_i$) of $V_i$, $a_{ij}$ denotes the causal coefficient from $V_j$ to $V_i$, and $\varepsilon_{\node{V}_i}$ represents the noise term.
Following this latent linear SCM setup, we generate multiple mock datasets with a random DAG containing both observed variables and latent variables, $\varepsilon_{\node{V}_i}$, randomly sampled from Gaussian distributions with a random mean and standard deviation $N(\mu_i,\sigma_i)$, and a random value of $a_{ij}$.
We apply the FCI algorithm to only observed variables, and we find the FCI algorithm is able to uncover the correct PAG corresponding to the ground-truth DAG in all of the generated datasets.

\section{Results}\label{sec:results}

\begin{figure}
    \centering
    \includegraphics[width=0.55\linewidth]{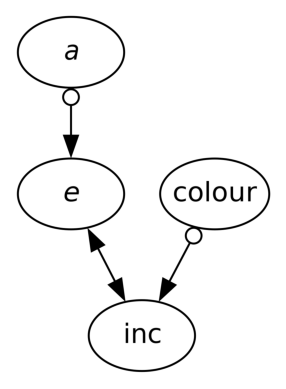}
    \caption{
    Partial Ancestral Graph (PAG) calculated with the Fast Causal Inference (FCI) algorithm \citep{SpirtesManuscript-SPIAAA,Spirtes:2013,zheng2024causal} with linear Fisher-Z conditional independence tests \citep{fisher1921probable} as well as non-linear Kernel-based conditional independence (KCI) tests \citep{Zhang:2012}.
    This PAG has three causal edges, which can be described as follows: (i) eccentricity is not an ancestor of the semimajor axis, (ii) there is a latent common cause of eccentricity and inclination, and (iii) inclination is not an ancestor of colour.
    }
    \label{fig:fci}
\end{figure}

Our primary finding is the causal structure found among $a$, $e$, inc, and colour, shown as a PAG in Fig.~\ref{fig:fci}.
The notations of the PAG can be found in \S\ref{sec:pag_notations}, and the PAG is derived through the FCI algorithm outlined in \S\ref{sec:fci_steps} and detailed in \S\ref{sec:steps}.
The PAG is based on the result of a set of CIs shown in \S\ref{sec:CI_tests} and Table~\ref{tab:CI}.
These CIs have confidence levels of at least 98.2\% and 91.5\% for Col-OSSOS data according to two different CI tests, and confidence levels of at least 99.9\% and 99.7\% for DES with the two CI tests.
Moreover, we consistently reproduce the same PAG as in Fig.~\ref{fig:fci} by jackknifing our data by sequentially leaving out each subpopulation of TNOs.
Thus, removing any subsample of 48 Classicals, 102 Resonant, 36 Centaurs, 28 Scattered, or 15 Detached TNOs among the Col-OSSOS data results in no change to our discovered PAG.
Therefore, we demonstrate that no single subpopulation is dominating the PAG and that our results are robust to outliers.

Alternatively, if we are to generate PAGs for the individual populations separately (i.e., analysing only one subpopulation at a time), we find a large diversity in the results.
Many of these PAGs, however, are based on very few data points.
Taking this result at face value hints that our overall PAG represents that main-line dynamics dominate over the entire sample.
We emphasize that this PAG was obtained with a purely data-driven approach, without astrophysical foresight.

We get a similar result when following the same procedure for the DES sample, except when removing the detached objects population where we end up with a slightly different PAG. 
However, running the model on the detached objects alone also leads to the same PAG, implying that while they contribute significantly to the result, they are not solely responsible for it as a subpopulation. 
The main difference in CIs that leads to this different PAG is the independence between $e$ and inc both with and without detached objects.
As the detached objects contribute to the high-eccentricity--high-inclination subpopulation while scattered objects have high eccentricity and low inclination, removing detached objects dilutes the global correlation between $e$ and inc. 
The detached subpopulation itself moreover does not have a strong correlation between the two since they do not undergo significant von Zeipel-Lidov-Kozai oscillations \citep{Zeipel:1910,Lidov:1962,Kozai:1962}.

\subsection{Astrophysical interpretation}

In this subsection, we investigate whether the causal links found by our model are consistent with the physical mechanisms at play in the Kuiper belt.
We emphasise that that our method is not capable of re-discovering these mechanisms or deriving physical laws from data, but should simply be compatible with them.

\subsubsection{$a\ \circ\hspace{-1.2mm}\rightarrow e$}

The first link we investigate is the one-way causal direction of the \textit{current} semimajor axis causing the \textit{current} eccentricity.
While the correlation between $a$ and $e$ in TNOs is well established, the direction of the causality we find here is not surprising either, as its root physical causes are:
\begin{itemize}
    \item Scattering by Neptune, where objects have to close-encounter Neptune first in order to get scattered into high-eccentricity orbits. 
    Moreover, objects usually cannot be both close to Neptune (today) and have a high eccentricity.
    It is the current semimajor axis of the objects that dictates what eccentricity they can have, and not the other way around. 
    \item Mean motion resonances (MMRs), where the period (and thus current semimajor axis) of the objects dictates whether they are inside an eccentricity-raising resonance. 
\end{itemize}
The connection $a\ \circ\hspace{-1.2mm}\rightarrow e$ \emph{rules out} the possibility of $a\leftarrow e$.
Clearly, $a\rightarrow e$ is possible, but also $a\leftrightarrow e$.
The latter might imply that an unobserved confounder causes both $a$ and $e$.

\subsubsection{$e\leftrightarrow$ inc}

The second link in the PAG is the two-way dependency between the eccentricity and the inclination, which is consistent with the von Zeipel-Lidov-Kozai anti-correlated oscillations between these two quantities (both inside and outside of MMRs), that plays a central role in the dynamics of TNOs.
Here, $e\leftrightarrow$ inc implies that there is an unobserved confounder.
Indeed, the von Zeipel-Lidov-Kozai mechanism involves perturbations from a third body, here being Neptune.
Moreover, if Neptune had not already been predicted in 1821 and eventually identified in 1846, our result here would strongly suggest the presence of an unknown perturbing body.
Together, the first two links successfully re-establish the main dynamical processes shaping the Kuiper belt (scattering, MMRs, and von Zeipel-Lidov-Kozai oscillations) without any physical inputs.  

\subsubsection{$\textrm{colour}\ \circ\hspace{-1.2mm}\rightarrow$ inc}

Finally, the third piece of the puzzle is the connection $\textrm{colour}\ \circ\hspace{-1.2mm}\rightarrow$ inc, \emph{ruling out} the possibility of $\textrm{colour}\leftarrow$ inc.
The `colour' (i.e., a proxy for the formation location in our null hypothesis) is hence consistent with causing the inclination ($\textrm{colour}\rightarrow$ inc).
This is again dynamically expected, as the formation location relative to inclination-raising secular resonances, such as f$_7$ and f$_8$, will strongly affect the inclination distribution of TNOs \citep{orangebook}.
However, this link leaves open the possibility of an unobserved confounder causing both colour and the inclination (colour $\leftrightarrow$ inc).
This confounder can be the formation location itself, if we were to assume the colour and initial location to be two distinct variables instead of the colour being a proxy for location. 

Our result, that $\textrm{colour}\leftarrow$ inc is not allowed, rules out the model of \citet{Luu:1996b} and \citet{Stern:2002}, where collisional evolution shapes the colours of TNOs.
Moreover, our result that $\textrm{colour}\leftarrow a$ is not allowed either, rules out models based solely on \citet{Zuzana}, where the current $a$ would control the amount of irradiation a TNO is subjected to, and thus its colour.
However, our findings cannot confirm or exclude the pre-instability irradiation scenario proposed by \citet{brown1} and \citet{wongbrown1,wongbrown}, as our analysis cannot disentangle the effect of formation location from immediate surface modification at the location of formation.
These findings are consistent with the recent results of \citet{belyakov} and \citet{licandro}.

\subsubsection{Further interesting features found in the PAG}

\begin{itemize}
    \item \emph{The lack of correlation between the colour and semimajor axis}: 
    This is dynamically expected as all TNOs in our sample underwent dynamical interactions with Neptune, that tend to be chaotic in nature.
    For example, many of the relevant processes (scattering, resonances, etc.) depend on the phase angle at which the TNO encounters Neptune.
    Some examples of the chaotic outcomes of the TNO dynamics are shown in \citet[][Figs.~11 and 12]{Ali-Dib:2021}.
    See also \citet[][Fig.~3]{nesvorny}.
    \item \emph{The indirect causation between the colour (initial location) and the eccentricity via the inclination}:
    Taken at face value, this would indicate that while the initial location directly causes the inclination, it is the final semimajor axis that causes the eccentricity.
    The effect of the initial semimajor axis on the eccentricity is indirect, and happens through von Zeipel-Lidov-Kozai oscillations starting from high inclinations.
\end{itemize}

\section{Discussion \& conclusions}\label{sec:discussion}

Our work endeavours to resolve the tension between theories of primordial origins vs.\ subsequent evolution to account for the observed dispersion and correlations in TNO colours, a subject of a long debate. 
Our causal graph analysis, derived from a model-agnostic causal discovery framework, strongly favours the primordial origin hypothesis that \emph{TNO colour is causally antecedent to inclination, not a consequence of it.}
Accordingly, we have high confidence (>91.5\%--99.9\%) in our result because it is unanimously found from both the Col-OSSOS and DES samples, with each dataset analysed by Fisher-Z and KCI tests.
While impacts undoubtedly modify surfaces, our results suggest they are not the dominant driver of colour diversity.
Moreover, our model seems to exclude any effects from the current semimajor axis on the colour of TNOs, disfavouring models where continuous irradiation plays a large role in shaping the colours. 
The consequences of having a colour gradient in the outer protoplanetary disk, as implied by this work (and many others as discussed), open up new possibilities to resolve the the Trojan colour conundrum \citep{jewitt}. 
This will be explored further in future work.

While many earlier works tried to explain the inclination--colour and eccentricity--colour correlations, both separately and simultaneously, our causal approach isolates inclination as the key dynamical variable causally linked to colour.
This hints at a larger role for inclination-raising secular resonances in the very early Solar System.
Indeed, \citet{Ali-Dib:2021} proposed that the origins of the paucity of VROs in the scattered disk are strongly linked to the f$_7$ and f$_8$ inclination modes.
In this scenario, the colour--eccentricity correlation is largely (although not necessarily entirely) a consequence of the more fundamental inclination--colour correlation, where the two can be linked via the von Zeipel-Lidov-Kozai mechanism.
This is consistent with the numerical model of  \citet{Ali-Dib:2021}, who proposed von Zeipel-Lidov-Kozai oscillations as a transport vehicle for VROs between high-inclination and high-eccentricity regimes.

Finally, our work is a proof of principle for the use of causality models in planetary sciences.
Moreover, our results will be a valuable addition to the study of Kuiper Belt Object colours with new data coming from the Vera C.\ Rubin Observatory.

\section*{Acknowledgements}
We thank the anonymous referee for their insightful comments that helped improve the clarity and quality of this manuscript.
This material is based on work supported by Tamkeen under the NYU Abu Dhabi Research Institute grant CASS.
YZ and KZ are supported by NSF Award No.~2229881, AI Institute for Societal Decision Making, NIH R01HL159805, and grants from Quris AI, Florin Court Capital, and MBZUAI-WIS Joint Program.
This research was carried out on the high-performance computing resources at New York University Abu Dhabi.
This research has made use of NASA's Astrophysics Data System Bibliographic Services.

%%%%%%%%%%%%%%%%%%%%%%%%%%%%%%%%%%%%%%%%%%%%%%%%%%
\section*{Data availability}
 
The data and code used for this work are available for download from the following GitHub repository: \href{https://github.com/ZehaoJin/causalTNOs}{\faGithub~\url{https://github.com/ZehaoJin/causalTNOs}}.

\section*{ORCID iDs}

\begin{CJK*}{UTF8}{gbsn}
\begin{flushleft}
Benjamin L.\ Davis \scalerel*{\includegraphics{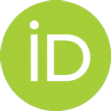}}{B} \url{https://orcid.org/0000-0002-4306-5950}\\
Mohamad Ali-Dib \scalerel*{\includegraphics{orcid-ID.pdf}}{B} \url{https://orcid.org/0000-0002-6633-376X}\\
Yujia Zheng (郑雨嘉) \scalerel*{\includegraphics{orcid-ID.pdf}}{B} \url{https://orcid.org/0009-0003-5225-6366}\\
Zehao Jin (金泽灏) \scalerel*{\includegraphics{orcid-ID.pdf}}{B} \url{https://orcid.org/0009-0000-2506-6645}\\
Kun Zhang (张坤) \scalerel*{\includegraphics{orcid-ID.pdf}}{B} \url{https://orcid.org/0000-0002-0738-9958}\\
Andrea Valerio Macci\`{o} \scalerel*{\includegraphics{orcid-ID.pdf}}{B} \url{https://orcid.org/0000-0002-8171-6507}
\end{flushleft}
\end{CJK*}

%%%%%%%%%%%%%%%%%%%% REFERENCES %%%%%%%%%%%%%%%%%%

% The best way to enter references is to use BibTeX:

\bibliographystyle{mnras}
\bibliography{example} % if your bibtex file is called example.bib

% Alternatively you could enter them by hand, like this:
% This method is tedious and prone to error if you have lots of references
%\begin{thebibliography}{99}
%\bibitem[\protect\citeauthoryear{Author}{2012}]{Author2012}
%Author A.~N., 2013, Journal of Improbable Astronomy, 1, 1
%\bibitem[\protect\citeauthoryear{Others}{2013}]{Others2013}
%Others S., 2012, Journal of Interesting Stuff, 17, 198
%\end{thebibliography}

%%%%%%%%%%%%%%%%%%%%%%%%%%%%%%%%%%%%%%%%%%%%%%%%%%

%%%%%%%%%%%%%%%%% APPENDICES %%%%%%%%%%%%%%%%%%%%%

% \appendix

%%%%%%%%%%%%%%%%%%%%%%%%%%%%%%%%%%%%%%%%%%%%%%%%%%

% Don't change these lines
\bsp	% typesetting comment
\label{lastpage}
\end{document}